\documentclass[12pt,a4paper]{article}
\usepackage{epsfig}
\def\eq#1{{Eq.~(\ref{#1})}}
\def\fig#1{{Fig.~\ref{#1}}}

\newcommand{\beq}{\begin{equation}}
\newcommand{\eeq}{\end{equation}}
\newcommand{\beqar}[1]{\begin{eqnarray}\label{#1}}
\newcommand{\eeqar}{\end{eqnarray}}
\newcommand{\ben}{\begin{eqnarray*}}
\newcommand{\een}{\end{eqnarray*}}

\newcommand{\as}{\alpha_s}
\newcommand{\un}{\underline}
\newcommand{\f}{\varphi}
\newcommand{\stackeven}[2]{{{}_{\displaystyle{#1}}\atop\displaystyle{#2}}}
\newcommand{\lsim}{\stackeven{<}{\sim}}

\begin{document}
\title {{\bf Elliptic Flow from Minijet Production~\vspace*{3mm} in Heavy
    Ion Collisions\\[1cm] }}
\author{
{\bf Yuri V. Kovchegov\thanks{e-mail: yuri@phys.washington.edu}~$\,
  ^a$ and
\quad  Kirill L. Tuchin\thanks{e-mail: tuchin@phys.washington.edu}~$\, ^b$
} \\[10mm]
{\it\small $^a$Department of Physics, University of Washington, Box 
351560}\\
{\it\small Seattle, WA 98195}\\[0.5cm]
{\it\small $^b$Institute for Nuclear Theory, University of 
Washington, Box 351550}\\
{\it\small Seattle, WA 98195}
}

\date{March, 2002}
\maketitle 
\thispagestyle{empty}

\begin{abstract}
We calculate the contribution to the elliptic flow observable $v_2$
from two--particle correlations in minijet production in
ultrarelativistic heavy ion collisions. We use a minijet production
cross section derived in a model inspired by saturation approach to
high energy scattering. Resulting differential elliptic flow $v_2
(p_T)$ is an increasing function of $p_T$ for transverse momenta below
the saturation scale $Q_s$. At higher transverse momenta ($p_T \, > \,
Q_s$) differential flow stops growing and becomes approximately
constant, reproducing the elliptic flow saturation data reported by
STAR. The centrality dependence of the minijet contribution to $v_2$
is also in good agreement with the data.
\end{abstract}
\thispagestyle{empty}
\begin{flushright}
\vspace{-19cm}
NT@UW--02--005\\
INT--PUB--02--32\\
\end{flushright}

\newpage

\setcounter{page}{1}
%
\section{Introduction}

Ultrarelativistic heavy ion collisions provide us interesting new
information on QCD under extreme conditions. Different stages of the
collisions probe QCD in different regimes: early times are likely to
be characterized by strong saturated gluonic fields \cite{glrmq,mv}
while creation of quark-gluon plasma (QGP) is an attractive
possibility in the later stages of the collisions \cite{qgp}. To be
able to understand the experimental data generated in the collisions
one has to learn to disentangle between the contributions of these two
(in principle) different physical processes to heavy ion observables.

The strong quasi-classical gluon fields are produced in the early
stages of the collisions due to high transverse densities of color
charges in the wave functions of the Lorentz-contracted nuclei
\cite{mv,yuri,jklw,claa,kv,yuaa}. The typical color charge density is
characterized by the {\it saturation} scale $Q_s^2$, which grows with
nuclear atomic number and with energy of the collision
\cite{glrmq,mv,bk,satevol,jimwlk}.  While being coherent over large
longitudinal distances these quasi-classical fields have a rather
short coherence length in the transverse direction, of the order of
the inverse saturation scale $1/Q_s$. Saturation-inspired models have
been quite successful in describing the emerging RHIC data on rapidity
and centrality dependence of the total charged multiplicity of the
produced particles \cite{dn,dgn}.

In the later stages of the collisions gluons and quarks generated by
the quasi-classical fields are likely to reach kinetic, and, possibly,
chemical thermal equilibrium, producing the quark-gluon plasma (QGP)
\cite{qgp,muller,lattice,bmss}. To be able to study this new state of
matter one should be able to distinguish the contributions to physical
observables of the collective phenomena due to QGP from the effects of
strong gluonic fields of the early stages. Ideally one should try to
construct observables which would be sensitive only to one type of
physics while being independent of the other. An example of such
observables are the long range rapidity correlations predicted in
\cite{klm}, which are almost entirely due to the dynamics of initial 
conditions.

In this paper we would like to study the contribution of the
quasi-classical gluonic fields to the elliptic flow observable $v_2$,
defined as the second Fourier moment of the azimuthal momentum
distribution of the produced particles \cite{elfl}. Elliptic flow
reflects anisotropy of the transverse momentum distribution of the
produced particles with respect to reaction plane. There has been a
large amount of elliptic flow data produced in the heavy ion
collisions at SPS \cite{na49,wa98} and RHIC
\cite{pho,star,phenix}. While the majority of these data are in good 
agreement with hydrodynamic simulations \cite{huov,teaney}, the
emerging new STAR data on differential elliptic flow $v_2 (p_T)$ at
high $p_T$ seems to deviate from hydrodynamic predictions
\cite{starsn}. Instead of continuing to increase with $p_T$ the flow variable
$v_2 (p_T)$ saturates to approximately a constant above $p_T \, = \,
1.5 \, \mbox{GeV}$ \cite{starsn}. The data goes up to $p_T \, \approx
\, 4.5 \, \mbox{GeV}$ \cite{starsn}. It is natural to expect that at such
high momenta the hard (perturbative) physics should be responsible for
the underlying strong interactions dynamics\footnote{Very recently an
attempt was made in \cite{tv} to explain the high-$p_T$ elliptic flow
using pure classical gluon fields of the nuclei \cite{claa} yielding
the elliptic flow which is too small to explain the data.}. This led
the authors of \cite{gvw} to propose a model of interplay of soft and
hard interactions in $v_2$: while the low momentum part of $v_2$ was
still described by hydrodynamic calculations the high $p_T$ part was
described by medium-induced radiative energy loss of partons
\cite{glvbdmps} which would generate azimuthal anisotropy in momentum
space due to coordinate space anisotropy of the overlap region. The
resulting flow observable would saturate at some relatively large
$p_T$ and then would decrease with increasing $p_T$. 

Here we are going to propose a model of a non-flow particle production
mechanism generating an elliptic flow observable $v_2$ which would go
to a constant at large $p_T$ and would stay approximately constant as
$p_T$ increases. Let us picture particle production at the early
stages of a nuclear collision at high energy. In the first
approximation high-$p_T$ particles are produced independent of each
other. We can illustrate this in the framework of the saturation
approach to nuclear collisions: there the gluon fields are coherent
over very short transverse distances of the order of $1/Q_s$, with
$Q_s \, \gg \,
\Lambda_{QCD}$. The high energy wave function of a large nucleus can
be viewed in the transverse plane as consisting of many independent
gluon fields each of them occupying a transverse area $1/Q_s^2$. (In
the longitudinal direction each classical field is of course coherent
over the whole nucleus \cite{mv} and this way the nucleus can be
considered as ``sliced'' into narrow pipes with diameter $1/Q_s$.) 
Therefore a collision with the overlap transverse area of the two
nuclei $S_\perp$ could be pictured as at least $S_\perp Q_s^2$
independent (sub)collisions. For sufficiently central heavy ion
collisions $S_\perp Q_s^2 \, \sim \, N_{part} \, \gg
\, 1$, where $N_{part}$ is the number of nucleons participating in the
collision. At the leading order in this large parameter $S_\perp
Q_s^2$ the two-particle multiplicity distribution factorizes into a
product of two single-particle multiplicity distributions. Correlation
between a pair of particles in the perturbative production mechanism
happens when both particles are produced in the same subcollision,
i.e., at the same impact parameter. Therefore these correlations
appear at the subleading in $S_\perp Q_s^2$ order and are suppressed
by a power of $S_\perp Q_s^2$. We are going to argue below that this
does not prevent them from significantly contributing to elliptic flow
observable.

The standard definition of differential elliptic flow is
\cite{vz,pv,ollie1,ollie2,bdopv}
\beq\label{v21}
v_2 (p_T) \, = \, \left< \, e^{ 2 i (\phi_{p_T} - \Phi_R)} \, \right> \, = \,
\left< \, \cos 2 (\phi_{p_T} - \Phi_R) \, \right>
\eeq
where $\phi_{p_T}$ is the azimuthal angle of the produced particle
with the value of transverse momentum $p_T$, $\Phi_R$ is the azimuthal
angle of the reaction plane and the brackets denote statistical
averaging over different events which leaves non-zero only the
contribution of the cosine in \eq{v21}. The reaction plane is
determined by averaging over particles produced in a heavy ion
collision with certain weights designed to optimize the reaction plane
resolution \cite{pv}. Let us imagine that the particle with momentum
$p_T$ in \eq{v21} was produced in the same subcollision with some
other particle, which contributed to determination of the reaction
plane angle $\Phi_R$. Then the particle $p_T$ would be correlated with
the reaction plane by these non-flow azimuthal correlations. The
averaged over events correlations do not disappear at high $p_T$, as
will be shown below. Therefore the elliptic flow observable $v_2$ may
be potentially sensitive to the non-flow correlations originating from
minijet production in the initial conditions. To avoid this problem
the authors of \cite{ollie1} (see also \cite{wang}) introduced a
cumulant approach to flow analysis. In terms of the saturation model
contribution of minijets to higher order cumulants defined in
\cite{ollie1,wang} is suppressed by powers of $S_\perp Q_s^2 \,
\sim \, N_{part}$. This led the authors of
\cite{ollie1,wang} to suggest that the high-cumulant flow analysis would 
be relatively free of minijet effects compared to standard flow
analysis.

There are certain potential dangers in this conclusion which have to
be addressed in an explicit calculation. One is that for peripheral
collisions $S_\perp Q_s^2 \, \sim \, N_{part}$ is not such a large
number and therefore minijet effects should not be suppressed
anymore. From the experimental data \cite{pho,star,phenix} we know
that elliptic flow is strongest in peripheral collisions making the
minijet contribution also large. Another potentially dangerous
question is whether the parameter $S_\perp Q_s^2$ appears in actual
calculations with some numerically small prefactor making inverse
powers of $S_\perp Q_s^2$ not too small. After all the observed
differential $v_2$ is also not a very large quantity being of the
order of $10-15 \%$ at RHIC and may be very sensitive to such
corrections.

To clarify the questions mentioned above we are going to perform an
explicit calculation of the minijet contribution to the elliptic flow
observable. The paper is structured as follows. In Sect. 2 we show
that the standard determination of flow employing \eq{v22} and the
flow extracted from the 2nd cumulant proposed in
\cite{ollie1,ollie2,wang} should yield the same result for $v_2$ for a
trivial choice of weights. This result was confirmed by an explicit
flow analysis using two different methods at STAR giving almost the
same result for $v_2$ \cite{tang}.

The second order cumulant from \cite{ollie1,wang} is nothing but a
two-particle correlation function, a contribution to which from the
minijet production is calculated in Sect. 3. To construct a model of
minijet production which incorporates both hard and soft physics we
have used $k_T$ factorized expression for two-jet production with the
unintegrated gluon distributions given by the quasi-classical
Glauber-Mueller expression \cite{Mue,meM,kt,coll,coll2,jklw}. These
gluon distributions are characterized by the saturation scale $Q_s \,
\gg \, \Lambda_{QCD}$ which insures applicability of small coupling
approaches down to rather small transverse momenta of the produced
particles. Our model of course does not yield us the exact
two-particle production cross section, but gives a realistic
approximation similar to the one used in
\cite{dn,dgn} to describe the multiplicity distributions at RHIC. An exact 
calculation of double inclusive minijet production in the
quasi-classical framework appears to be rather complicated and is left
for further research: even the single inclusive cross section has not
been yet unambiguously theoretically determined, despite the extensive
efforts \cite{kv,yuaa}. 

In Sect. 3 we calculate the differential flow observable resulting
from these two-particle correlations. Our final result is given in
\eq{V2}. The obtained differential elliptic flow $v_2 (p_T)$ starts 
increasing as a power of $p_T$ for small $p_T \, \ll \, Q_s$ (see
\eq{small}) and then saturates to a slow logarithmic growth for $p_T \,
\gg \, Q_s$ (see \eq{large}). We also derive the centrality dependence 
of the minijet flow contribution (see \eq{bdep}). 

In Sect. 4 after making some simple assumptions about the gluon
distributions employed in the minijet production cross section we fit
the differential elliptic flow STAR data \cite{starsn} using the flow
from \eq{V2} with $\as \, = \, 0.3$, $Q_s \, = \, 1 \,
\mbox{GeV}^2$. These values are in agreement with the
ones used in the saturation-inspired analysis of the multiplicity data
in \cite{dn,dgn}. The fit is shown in \fig{ptfig}.  We can also fit
the centrality dependence of $v_2$ with our minijet model by noting
that the integrated flow should scale as $v_2 (B) \, \sim \,
1/\sqrt{S_\perp Q_s^2} \, \sim \, 1/\sqrt{N_{part}}$ (see
\fig{btfig}). 

We discuss the results in Sect. 5 by stating that while a more
involved numerical analysis is still needed to analyze the emerging
RHIC data on elliptic flow we have demonstrated that the contribution
of minijets to the standard flow analysis is very large, possibly
accounting for most of high-$p_T$ data. Thus it appears that the
standard flow analysis is heavily ``contaminated'' by minijets which
prevent direct measurements of the contribution of collective QGP
effects to elliptic flow. At the same time flow analysis seems to be
rather sensitive to details of saturation physics and could be used
for determination of nuclear saturation scales.

\section{Different Methods of Flow Analysis}

In \cite{ollie1} Borghini et al proposed a new and interesting
approach to flow analysis employing higher order cumulants. Let us
outline some important features of the approach presented in
\cite{ollie1} for the lowest order cumulant. The second order
cumulant is defined for two particles with azimuthal angles $\phi_1$
and $\phi_2$ as \cite{ollie1,ollie2,wang}
\beq\label{c21}
\left<\left< \, e^{2 i (\phi_1 - \phi_2)} \, \right>\right> \, = \, 
\left< \, e^{2 i (\phi_1 - \phi_2)} \, \right> - \left< \, 
e^{2 i \phi_1} \, \right> \, \left< \, e^{- 2 i \phi_2} \, \right>.
\eeq
The average $\left< \, e^{2 i \phi_1} \, \right>$ vanishes since the
angle here is measured in the laboratory and heavy ion collisions are
azimuthally symmetric after averaging over many events. For flow
correlations this means that the angle in $\left< \, e^{2 i \phi_1} \,
\right>$ is not measured with respect to reaction plane, but with respect 
to some fixed direction in the detector. Assuming that the particles
$1$ and $2$ are correlated with each other only through the flow
correlations with the reaction plane the authors of \cite{ollie2}
wrote
\beq\label{c22}
\left<\left< \, e^{2 i (\phi_1 - \phi_2)} \, \right>\right> \, = \, 
\left< \, e^{2 i (\phi_1 - \phi_2)} \, \right> \, = \, 
\left< \, e^{2 i (\phi_1 - \Phi_R)} \, e^{2 i (\Phi_R - \phi_2)}\, \right> \,
= \, (v_2)^2, 
\eeq
where the definition of elliptic flow from
\eq{v21} was employed. A new method for measuring elliptic 
flow was proposed in \cite{ollie1} using the two particle (and higher
order) cumulants of \eq{c22}. If we fix the transverse momentum of
particle $1$ to be $p_T$ and average over all momenta of the particle
$2$ over many events then as one can see from \eq{c22}
\beq\label{c23}
\left< \, e^{2 i (\phi_1 (p_T) - \phi_2)} \, \right> \, = \, v_2 (p_T) \left< 
v_2 \right>
\eeq
with $\left< v_2 \right>$ the elliptic flow variable averaged over all
$p_T$. At the same time if we do not impose any restrictions on the
transverse momenta of both particles we get \cite{ollie2}
\beq\label{c24}
\left< \, e^{2 i (\phi_1 - \phi_2)} \, \right> \, = \, \left< 
v_2 \right>^2.
\eeq
From Eqs. (\ref{c23}) and (\ref{c24}) noting that only cosine
components of the exponents survive the averaging we obtain the
following expression for differential elliptic flow \cite{ollie2}
\beq\label{v2o}
v_2 (p_T) \, = \, \frac{\left< \, \cos ( 2 (\phi_1 (p_T) - \phi_2)) \,
\right>}{\sqrt{\left< \, \cos (2 (\phi_1 - \phi_2)) \, \right>}} .
\eeq
Let us define the event-averaged two-particle multiplicity
distribution function
\beq\label{2pd}
P(k_1, k_2, {\un B}) \, = \, \frac{dN}{d^2 k_1 \, dy_1 \, d^2 k_2
\,dy_2} ({\un B}),
\eeq
where the transverse momenta of the particles are ${\un k}_1$ and
${\un k}_2$, while $y_1$ and $y_2$ are their rapidities and we average
over all events with the impact parameter ${\un B}$ between the two
nuclei. This distribution can be written as a sum of uncorrelated and
correlated terms
\beq\label{COR}
P( k_1, k_2, {\un B})\,=\,
\frac{dN}{d^2 k_1\, dy_1}\frac{dN}{d^2 k_2\, dy_2} + 
\frac{dN_{corr}}{d^2 k_1\,dy_1\, d^2 k_2\,dy_2},
\eeq
where we suppressed the impact parameter dependence. The correlated
term in \eq{COR} is usually much smaller than the uncorrelated one as
it is suppressed by a power of $Q_s^2 S_\perp \, \sim \,
N_{part}$. Using \eq{2pd} we rewrite \eq{v2o} as
\ben
v_2(k_1, {\un B})=\frac{\int\, d^2 k_2 dy_2 \, d\phi_1 dy_1\,P( k_1,
k_2, {\un B})\,
\cos(2(\phi_1-\phi_2))} {\int\, d^2 k_2 dy_2\, d\phi_1 dy_1\,
P(k_1, k_2, {\un B})} 
\een
\beq\label{MAIN}
\times \, \left(\frac{\int\, d^2 k_1 dy_1\, d^2 k_2 dy_2\, 
P(k_1\, k_2, {\un B})}{\int\, d^2 k_1 dy_1\, d^2 k_2 dy_2\, P(k_1,
k_2, {\un B})
\cos(2(\phi_1-\phi_2))} \right)^{1/2} ,
\eeq
where we have relabeled the transverse momenta of the particles to be
${\un k}_1$ and ${\un k}_2$. \eq{MAIN} gives us a way of calculating
elliptic flow from two-particle correlation functions. Higher order
cumulants would yield us ways of calculating flow from higher order
particle correlations \cite{ollie1}, though one can not go to
arbitrary high order cumulants due to lack of statistics there. An
analysis of RHIC data has been performed at STAR \cite{tang} using
both conventional and cumulant approaches explicitly demonstrating
that the flow obtained by the second cumulant technique of \eq{MAIN}
is consistent with the flow extracted using conventional
techniques. That is the approach of \eq{MAIN} is equivalent to the
conventional flow analysis. To see this let us first note that in the
actual standard flow analysis one has to take into account the
resolution in the event plane determination \cite{pv}. This would
modify \eq{v21} giving \cite{pv,pho}
\beq\label{v22}
v_2 (p_T) \, = \, \frac{\left< \, \cos (2 (\phi_{p_T} - \Psi_R)) \,
\right>}{\sqrt{2 \left< \, \cos (2 (\Psi_R^a - \Psi_R^b)) \, \right>}}
\eeq
where $\Psi_R^a$ and $\Psi_R^b$ are the event plane angles determined
in two different sub-events labeled $a$ and $b$ with the particle
multiplicity $N/2$ in each of them while $\Psi_R$ is the event plane
angle determined in the full event with multiplicity $N$
\cite{pv}. \eq{v22} is quite similar to \eq{v2o}. To show that the two
equations are almost equivalent let us go back to exponential notation
and rewrite the numerator of \eq{v2o} as
\beq\label{cos1}
\left< \, \cos ( 2 (\phi_1 (p_T) - \phi_2)) \,
\right> \, = \, \left< \left< e^{2 i (\phi_1 (p_T) - 
\phi_2 (k_T))}\right>_{all \, k_T\neq p_T} 
\right>_{events}
\eeq
where we first average over all particles (with various $k_T$) in a
given event other than the chosen particle with momentum $p_T$, just
like in the standard flow analysis \cite{star}, and then average over
all events. On the other hand, the azimuthal angle of the reaction
plane $\Psi_R$ in \eq{v22} can be defined by
\beq\label{PR}
Q_N \, e^{- 2 i \Psi_R} \, = \, \left< e^{- 2 i \phi_{k_T}}
\right>_{all \, k_T\neq p_T},
\eeq
where $Q_N$ is a real number. If one neglects correlations between the
particles and multiplies \eq{PR} by its complex conjugate avergaing
over events with the same total multiplicity $N$ one easily gets $<
Q_N^2 > \, = \, 1/N$ \cite{ollie1}. Therefore at the leading order in
$N$ we have (for large $N$) $Q_N \, \sim \, 1/\sqrt{N}$.  In
\eq{PR} we for the moment forget about the transverse momentum cutoffs
imposed in event plane determination and subtleties related to
different choices of weights and detector imperfections. Substituting
\eq{PR} into the numerator of \eq{v22} we obtain
\ben
\left< \, \cos (2 (\phi_{p_T} - \Psi_R)) \,
\right>_{events} \, = \, \left< \, e^{2 i (\phi_{p_T} - \Psi_R)} \,
\right>_{events} \, = \, \left<  \frac{1}{Q_N} \, \left< \, 
e^{2 i (\phi_{p_T} - \phi_{k_T})}
\, \right>_{all \, k_T\neq p_T} \right>_{events} 
\een
\beq\label{cos2}
\approx \frac{1}{\left< Q_N \right>} \, \left< \left< \, 
e^{2 i (\phi_{p_T} - \phi_{k_T})}
\, \right>_{all \, k_T\neq p_T} \right>_{events} 
\eeq
where we assumed that for a fixed impact parameter collisions the
particle multiplicity $N$ (and, therefore, $Q_N$) is independent of
two-particle correlations. \eq{cos2} is identical to \eq{cos1} up to a
factor of $1/\left< Q_N \right>$. Similarly one can show that the
denominators in Eqs. (\ref{v22}) and (\ref{v2o}) are equal to each
other up to a factor of $\sqrt{2 / \left< Q_{N/2}^2 \right>}$:
\ben
2 \left< \, \cos (2 (\Psi_R^a - \Psi_R^b)) \, \right>_{events} \, = \,
2 \left< \, \frac{1}{Q_{N/2}^2} \, \left< e^{2 i (\phi_{k_{T1}} -
\phi_{k_{T2}})}\right>_{all \, different \, k_{T1}, k_{T2} } \,
\right>_{events}
\een
\beq\label{cos3}
\approx \, \frac{2}{\left< Q_{N/2}^2 \right>} \, 
\left< \cos (2 (\phi_1 - \phi_2))\right>. 
\eeq
Recalling that for large multiplicities $\left< Q_N \right> \sim
1/\sqrt{N}$ we see that at the leading order in $N$: $\sqrt{\left<
Q_{N/2}^2 \right>} / \sqrt{2} \left< Q_N \right> \, \approx \, \left<
Q_{N/2} \right> / \sqrt{2} \left< Q_N \right> \approx 1$. Extra
factors in the numerator and denominator of \eq{v22} cancel reducing
it to \eq{v2o}.  Thus we showed that Eqs. (\ref{v22}) and (\ref{v2o})
are identical in the limit of large multiplicity $N$. We have proven
that the standard definition of flow from \eq{v22} is equivalent to
the definition introduced in \cite{ollie1,wang} given here by
\eq{v2o}). Therefore in order to calculate the differential elliptic
flow $v_2 (p_T)$ we need only to calculate the two-particle
correlation function $P (k_1, k_2, {\un B})$ and substitute it into
\eq{MAIN}.

\section{Calculation of $v_2$}

To calculate the two-particle correlation coefficient in \eq{COR} one
needs to know single and double particle multiplicity
distributions. While the quasi-classical single particle production
mechanisms in heavy ion collisions have been extensively studied
\cite{claa,kv,yuaa}, the double inclusive particle production has not 
been calculated yet. One needs to calculate production of a pair of
particles with comparable transverse momenta $|{\un k}_1| \sim |{\un
k}_2| \sim Q_s$ and rapidities $y_1 \sim y_2$ in the framework of
McLerran-Venugopalan model \cite{kv}, that is resumming all powers of
$Q_s^2/k_\perp^2$ \cite{yuri,meM,yuaa}. Unlike the single gluon
production this process involves an extra gluon and can not be
described by the classical field methods of \cite{claa,kv,yuaa}. At
the lowest order in $Q_s^2/k_\perp^2$ the two gluon production
amplitude is equivalent to the real part of NLO BFKL kernel
\cite{BFKL,NLO} and is rather sophisticated \cite{leostr}. Going
beyond leading order in $Q_s^2/k_\perp^2$ appears to be tremendously
complicated and we will not address this problem here. Instead we are
going to construct a model of correlated two-particle production
employing quasi-classical gluon distributions from \cite{jklw,meM} in
the production formula inspired by collinear factorization
\cite{coll,coll2}, similar to how it was done in \cite{dn,dgn} in 
describing the multiplicity data at RHIC. We would also assume that in
any given event either $y_1 \gg y_2$ or $y_1 \ll y_2$. Of course this
assumption does not hold in actual flow analyses \cite{star,phenix,pho}
and we are making it just to simplify the calculations. While, as
discussed above, a more detailed calculation would still be required
to obtain an exact expression for the correlation coefficient in
\eq{COR}, we believe that it would only introduce numerical corrections 
to our approach leaving qualitative results the same.

\begin{figure}
\begin{center}
\begin{tabular}{cc}
\epsfig{file=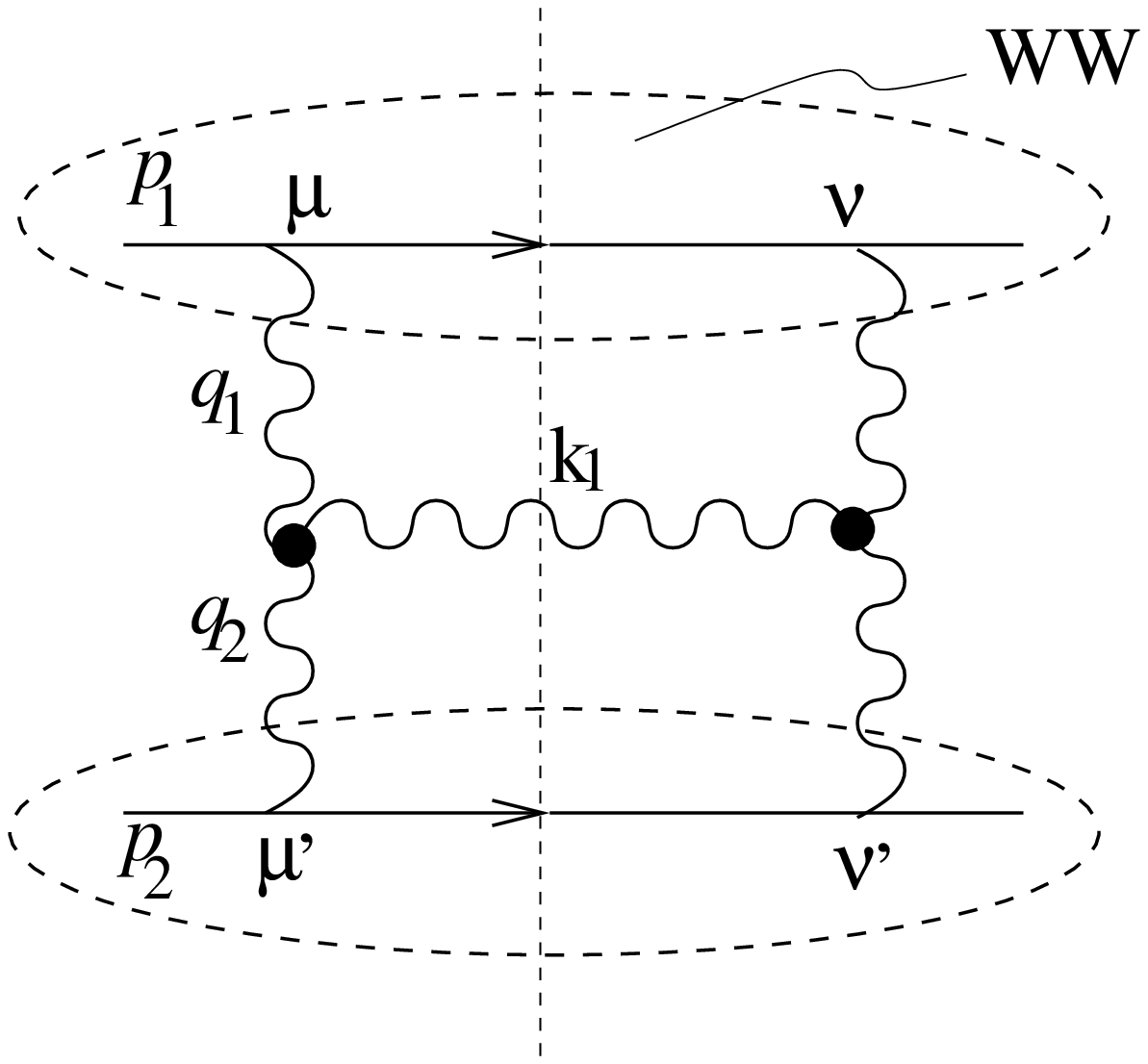, width=7cm}
&
\epsfig{file=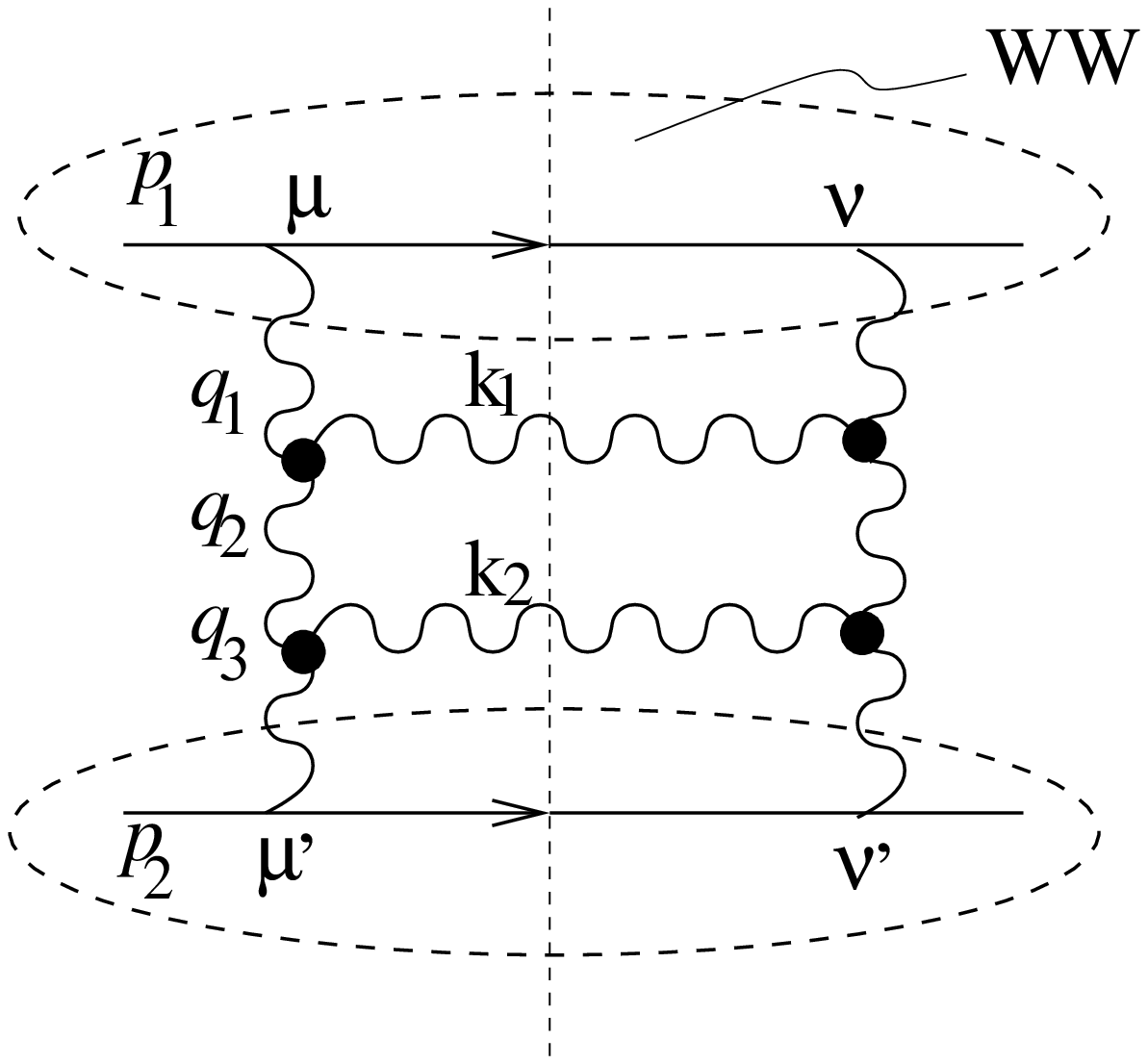, width=7cm}\\
{\sl (a)} & {\sl (b)}
\end{tabular}
\end{center}
\caption{\sl (a) One- and (b) two-gluon production amplitudes. Thick dots 
denote Lipatov vertices. }
\label{FPROD}
\end{figure}

Consider inclusive production of one or two gluons in scattering of
two quarks at high energy as shown in \fig{FPROD}, where the blobs
denote effective Lipatov vertices. It is convenient to use the Sudakov
decomposition of gluons' four-momenta:
\beq\label{SUDAKOV}
q_i\, =\, \alpha_i p_1\, +\, \beta_i p_2 \, +\, \un q_i, \quad i=1, 2,
3.
\eeq
In \fig{FPROD}(a) the dominant contribution stems from the multi-regge
kinematical region where $\alpha_1\gg\alpha_2$ and
$\beta_2\gg\beta_1$. Using Ward identities the $t$-channel gluon
propagators can be written as
\beq\label{PROPAG}
D^{\mu\nu}(q_i)\,=\,\frac{2}{s}\frac{\un q_i^\mu\un
  q_i^\nu}{\alpha_i\beta_i},
\eeq
where $s\,=\, 2(p_1\cdot p_2)$ is center-of-mass energy squared. With
this representation of the $t$-channel gluon propagator effective
Lipatov's vertex \cite{BFKL} for $s$-channel gluon production (black
blobs in the \fig{FPROD}) reduces to the usual three-gluon vertex
$\Gamma_{\mu\nu}^\rho$ and can be easily calculated:
\beq\label{LIPAT}
L^\rho\, =\, \un q_1^\mu \un q_2^\nu \Gamma_{\mu\nu}^\rho\, =\,
\frac{1}{2}\,\alpha_1\beta_2\,s\,
\left[\left(\alpha_1+\frac{2\un q_1^2}{\beta_2 s}\right)\, p_1^\rho\,
  +\,\left(\beta_2+\frac{2\un q_2^2}{\alpha_1 s}\right)\, p_2^\rho\,-
\, (\un q_1+\un q_2)^\rho\right].
\eeq
In our calculation we will need the square of Lipatov's vertex given
by
\beq\label{SQLIP}
(L^\rho)^2\,=\, \alpha_1\,\beta_2\,s \,\un q_1^2\un q_2^2.
\eeq
Using \eq{PROPAG}, \eq{LIPAT}, \eq{SQLIP} and the four-dimensional
momentum element decomposition 
\beq
d^4 q_i\,=\,\frac{s}{2}\, d^2\un q_i\, d\alpha_i\, d\beta_i
\eeq  
we evaluate the single and double distribution of the produced
particles shown in \fig{FPROD}
\beq\label{AMPL1}
\frac{dN}{d^2 k_1\,dy_1} \,=\, 
\frac{\alpha_s^3}{\pi^2}\,\frac{2 \, C_F \, A^2}{S_\perp \, {\un k}_1^2}\,
\int\, d^2 q_1 \,
\frac{1}{\un q_1^2 \, (\un k_1 -\un q_1)^2},
\eeq
\beq\label{AMPL2}
\frac{dN_{corr}}{d^2 k_1\,dy_1\, d^2 k_2\,dy_2} \,=\, 
\frac{\alpha_s^4}{\pi^4}\, \frac{N_c \, C_F \, A^2}{S_\perp \, 
{\un k}_1^2 \, {\un k}_2^2} \, 
\int\, d^2 q_1 \, \frac{1}{\un q_1^2 \, (\un k_1 + \un k_2 - \un q_1)^2},
\eeq
where we changed variables to $\un k_1\,=\, \un q_1-\un q_2$, $\un
k_2\,=\, \un q_2-\un q_3$ (the momenta of produced gluons).
Throughout the paper we assume for simplicity that the nucleus has a
cylindrical shape with radius $R$ and ``height'' $2 R$ and the axis of
the cylinder coincides with the collision axis. $S_\bot=S_\bot(\un B)$
is the nucleus overlap transverse area which depends on nuclei
relative impact parameter $\un B$. We assume that colliding nuclei are
identical having atomic number $A$ each.

In fact colliding quarks in \fig{FPROD} are not free but confined
inside nucleons. If momenta of produced gluons are higher than all
scales characterizing the nucleus (including $Q_s$) we can use
$k_t$-factorization to write the distributions from \eq{AMPL1} and
\eq{AMPL2} as convolutions of hard amplitudes with parton
distributions in nuclei which are predominantly gluonic in the high
energy region $s\gg t$. Employing the relation \cite{meM}
\beq\label{xG}
xG_A(x,\un q^2)\,=\,A\,xG(x,\un q^2)\,=\, 
A\,\frac{\alpha_s C_F}{\pi}\, \ln(\un q^2/\mu^2),
\eeq
with $\mu$ some non-perturbative cutoff the one and two-particle
distributions can be written as \cite{lr}
\beq\label{DISTR1}
\frac{dN}{d^2 k_1\,dy_1}\,=\,
\frac{2 \, \as\,}{C_F \, S_\bot}\,\frac{1}{\un k_1^2}\,\int\,
d^2 q_1\frac{dxG_A}{d\un q_1^2}\,\frac{dxG_A}{d(\un k_1- \un q_1)^2}
\eeq
\beq\label{DISTR2}
\frac{dN_{corr}}{d^2 k_1\,dy_1\, d^2 k_2\,dy_2}\,=\,
\frac{N_c \, \as^2}{\pi^2 \, C_F \, S_\bot}\,
\frac{1}{\un k_1^2\,\un k_2^2}\,\int\,
d^2 q_1\frac{d xG_A}{d\un q_1^2}\,\frac{d xG_A}{d(\un k_1+\un k_2 -
\un q_1)^2}.
\eeq

As was argued in the Introduction, if the number of partons in each
nuclei is large enough then their mutual interactions must be taken
into account. This may happen either due to increase of parton density
in each nucleon by subsequent emission of gluons in course of quantum
evolution as collision energy increases \cite{bk,jimwlk} or due to
enhancement of nucleon parton density in a large nucleus by the atomic
number $A$ \cite{mv,Mue,yuri}. The latter effect leads to creation of
the non-Abelian Weisz\"acker-Williams (WW) field $\un A^{WW}(\un z)$
of the nucleus which is a strong classical gluon field already at
moderate energies \cite{yuri,jklw}. The WW field gives rise to the
unintegrated gluon distribution given by \cite{jklw,meM}
\begin{eqnarray}
\frac{dxG_A(x,\un q^2)}{d\un q^2} &=&
\frac{2}{(2\pi)^2}\,\int\, d^2\un z \,e^{-i\un z\cdot\un q}\,
\int\, d^2\un b \,\mathrm{Tr}\,\langle \un A^{WW}(\un 0)\,\un A^{WW}(\un
z)\rangle\nonumber\\
&=&
\frac{2}{\pi (2\pi)^2}\,\int\, d^2\un z\, e^{-i\un z\cdot\un q}\,
\frac{S_\bot C_F}{\as\,\un z^2}\left(
1-e^{-\frac{1}{4}\un z^2 Q_s^2}\right),\label{WW}
\end{eqnarray}
where $\un b$ is the gluon's impact parameter (which we can trivially
integrate over in a cylindrical nucleus) and
\beq\label{SATSCALE}
Q_s^2(\un z)\,=\, \frac{4\pi^2\, \as N_c}{N_c^2 - 1}\,\rho\,
xG(x,1/\un z^2)\, T(\un b),
\eeq
with $\rho\,=\, A/[2 \pi R^3]$ the atomic number density in the
cylindrical nucleus with atomic number $A$, and $T(\un b)$ the nucleus
profile function equal to $2 R$ for the cylindrical nucleus considered
here.  This provides the initial condition to the nonlinear quantum
evolution of the gluon distribution with energy in the high parton
density region \cite{bk}. $Q_s$ is a scale at which nonlinear nature
of the gluon field becomes evident. We suggest using the classical
expression \eq{WW} as an approximation to the exact gluon field of the
nucleus. This is a justified approximation as long as $\as\ln s\,
\lsim \,1$, i.e.,\ when corrections due to quantum evolution are small.

It is usually assumed that $k_T$-factorization holds in high parton
density regime as well as in the linear one
\cite{kt}. Phenomenological models for heavy ion collisions which
employ this assumption together with nonlinear evolution for gluon
distributions proved to be successful in describing experimental data
at SPS and RHIC \cite{dgn}. This implies that the use of
$k_T$-factorization is a quite good approximation for the SPS and RHIC
kinematical regions even for $k_T < Q_s$. Therefore substituting
\eq{WW} into \eq{DISTR1} and into \eq{DISTR2} we obtain the following 
expressions for the single and double gluon distributions
\beq\label{DD1}
\frac{dN}{d^2 k_1\, dy_1}\,=\frac{C_F\, S_\bot}
{\as\,\un k_1^2}\,\frac{4 \, K_1}{\pi^3}\,
\int_0^\infty\,\frac{dz}{z^3}\, J_0(k_1z)\,\left(
1-e^{-\un z^2 Q_s^2/4}\right)^2,
\eeq
\beq\label{DD2}
\frac{dN_{corr}}{d^2 k_1\, dy_1 \, d^2 k_2\, dy_2}\,=\frac{N_c \, C_F \, S_\bot}
{\un k_1^2\,\un k_2^2}\,\frac{K_2}{\pi^{6}}\,
\int d^2 \un z \,\frac{1}{\un z^4} \,
e^{- i\un z \cdot (\un k_1+\un k_2)}\,\left(
1-e^{-\un z^2 Q_s^2/4}\right)^2,
\eeq
where the saturation scales in both nuclei are the same since the
nuclei are cylindrical and identical. Generalization of
Eqs. (\ref{DD1}) and (\ref{DD2}) to a spherical nucleus is
straightforward.

We have to point out again that the single particle distribution is,
in principle, known better than displayed here in \eq{DD1}
\cite{yuaa}. For instance the distribution in \eq{DD1} is not infrared
safe, while the correct distribution derived in \cite{yuaa} has no
infrared divergences.  However, as we need both single and double
particle distributions to obtain elliptic flow using \eq{MAIN} we
should calculate both of them in the framework of the same
model. Since the exact calculation of the double gluon distribution
does not seem feasible at the moment we have to calculate both
distributions in the same $k_T$-factorization approach inserting
$K$-factors $K_1$ and $K_2$ to correct the normalization of the
approximation to include higher order corrections \cite{dgn}, which we
have done in Eqs. (\ref{DD1}) and (\ref{DD2}). The value of the
$K$-factors will be fixed later. We will determine $K_1$ by comparing
particle multiplicity per unit rapidity ($dN/dy$) resulting from
\eq{DD1} to the total multiplicity observed at RHIC
\cite{dgn}. To fix $K_2$ we consider two-particle production cross 
section, which is proportional to the two-particle multiplicity
distribution function $P (k_1, k_2, {\un B})$ from \eq{COR}. In the
limit of large transverse momentum $k_1 \sim k_2 \sim p_T$ the second
term in \eq{COR} given by \eq{DD2} falls off at most as $1/p_T^6$ and
has a collinear singularity ${\un k}_1 + {\un k}_2 = 0$.  Therefore it
dominates over the first term in \eq{COR} given by \eq{DD1} squared,
which gives a $1/p_T^8$ fall off of the cross section. For large ${\un
k}_1 = - {\un k}_2$ \eq{COR} and, therefore, \eq{DD2} should match
onto the corresponding collinear factorization expression for
back-to-back jets \cite{coll,coll2,lr,coll3,hijing}. To fix the
normalization we expand the term in the parentheses of \eq{DD2} to the
lowest order and integrate over one of the transverse momenta
obtaining
\beq\label{scoll}
\frac{d \sigma_{corr}}{d p_T^2 \, d y_1 \, d y_2} \, = \, K_2 \, 
\frac{9 \, \pi \, \as^2}{4 \, p_T^4} \, [xG (x, p_T^2)]^2
\eeq
in agreement with collinear factorization result at mid-rapidity
\cite{coll,coll2,coll3,hijing}. We put $N_c = 3$ explicitly in \eq{scoll}. 
Collinear factorization models \cite{coll,coll2,coll3,hijing} are
rather successful in describing the high-$p_T$ particle spectra in
hadronic and heavy ion collisions when putting $K_2 = 2$ to correct
the lowest order perturbative expression for next-to-leading order
effects. For our model to be in agreement with these high-$p_T$ data
we will have to also put $K_2 = 2$ when trying to describe the flow
data in the next section. We have to note that since the gluon
distributions in \eq{WW} do not include DGLAP \cite{dglap} evolution
in them and \eq{DD2} does not have jet quenching effects
\cite{gvw,glvbdmps,hijing} in it our model can not be applied to
describe the high-$p_T$ spectra and a complete DGLAP-evolved gluon
distributions along with jet quenching should be used in a more
detailed numerical treatment of the problem as was done for particle
spectra in \cite{coll,coll2,coll3,hijing}.

Substituting Eqs. (\ref{DD1}) and (\ref{DD2}) into \eq{COR} we are
ready to calculate $v_2 (p_T, {\un B})$. To calculate the integral in
the numerator of the first line of \eq{MAIN} we note that only the
correlated part of $P(k_1, k_2, {\un B})$ (the second term in
\eq{COR}) contributes there. Let us denote the angle between
two-vectors $\un z$ and $\un k_1$ by $\theta$ and the angle between
$\un k_2$ and $\un k_1$ by $\phi_2$. The third angle defining direction
of $\un k_1$ is free, so we can just integrate over it obtaining a
factor of $2 \pi$. Plugging \eq{DD2} into \eq{COR} and using the
latter in \eq{MAIN} we get
\begin{eqnarray}
&& \int\, d^2 k_2\, d\phi_1\, P(k_1, k_2, {\un B}) \,
\cos(2(\phi_1-\phi_2)) \, = \, \int\, d^2 k_2\, d\phi_1\,
\frac{dN_{corr}}{d^2k_1\, dy_1\, d^2 k_2\, dy_2}\,
\cos(2(\phi_1-\phi_2))\nonumber\\
&&=\,\frac{N_c \, C_F \, S_\bot}{\un k_1^2}\,\frac{K_2}{\pi^5}\,
\int\, \frac{dz}{z^3}\, d\theta\, \frac{dk_2^2}{
  k_2^2}\,\left( 1-e^{-\un z^2 Q_s^2/4}\right)^2\, 
\int_0^{2\pi} d\phi_2\, e^{-izk_2\cos(\phi_2-\theta)} \,
e^{-izk_1\cos\theta}\, \cos2\phi_2\nonumber\\ &&= \frac{N_c \, C_F \,
S_\bot}{\un k_1^2}\,\frac{4 \, K_2}{\pi^3}\,\int\frac{dz}{z^3}\,
\frac{dk_2^2}{ k_2^2}\,
\left( 1-e^{-\un z^2 Q_s^2/4}\right)^2\, J_2(k_1z)\, J_2(k_2z)\nonumber\\
&&= \frac{N_c \, C_F \, S_\bot}{\un
k_1^2}\,\frac{4 \, K_2}{\pi^3}\,\int_0^\infty\frac{dz}{z^3}\,\left( 1-e^{-\un
z^2 Q_s^2/4}\right)^2\, J_2(k_1z)\label{I1}.
\end{eqnarray}
To calculate the integral in the denominator of the first line of
\eq{MAIN} we first note that the final state multiplicity per unit 
rapidity in heavy ion collisions was calculated in \cite{yuaa} (see
also \cite{Mueller2}) and reads
\beq\label{TOTMULT}
\frac{dN}{dy}\,=\,c \, \frac{S_\bot\, C_F\, Q_s^2}{\as\, 2\, \pi^2}
\eeq
with $c$ the gluon liberation coefficient which was calculated in
\cite{yuaa} to be $c = 2 \ln 2$. Following the same steps as in derivation of
\eq{I1} we find contributions to the integral in the denominator of
the first line of \eq{MAIN} of uncorrelated two-particle distribution
\begin{eqnarray}
&&\int\, d^2 k_2\, d\f_1\,
\frac{dN}{d^2 k_1 \, dy_1}\frac{dN}{d^2 k_2 \, dy_2}\,=\,
\frac{dN}{dy_2}\,\frac{dN}{k_1 \, d k_1\,dy_1} \nonumber\\ &&= 
\,c \, \frac{C_F^2\, S_\bot^2\, Q_s^2} {\as^2\,\un k_1^2}\,\frac{4 \, K_1}{\pi^4}\,
\int_0^\infty\,\frac{dz}{z^3}\, J_0(k_1z)\,\left(
1-e^{-\un z^2 Q_s^2/4}\right)^2
\label{I4}
\end{eqnarray}
and of correlated two-particle distribution
\begin{eqnarray}
&&\int\, d^2 k_2\, d\f_1\,\frac{dN_{corr}}{d^2 k_1\, dy_1\, d^2 k_2 \,
dy_2}\nonumber\\ &&= \frac{N_c\, C_F \, S_\bot}{\un k_1^2}\,\frac{8 \,
K_2}{\pi^3}\,\int_0^\infty\frac{dz}{z^3}\,\left( 1-e^{-\un z^2
Q_s^2/4}\right)^2\, J_0(k_1z)\, \ln\frac{1}{z\mu}.
\label{I2}
\end{eqnarray}
As one can obviously see the integral in \eq{I4} is enhanced by factor
of $S_\bot Q_s^2/\as^2$ as compared to the integral in \eq{I2}. This
means that the total number of correlated pairs of particles is
negligible compared to the number of the uncorrelated pairs given by
the square of the total multiplicity. Therefore the correlated
contribution form \eq{I2} can be neglected compared to the
uncorrelated contribution in \eq{I4}.

\begin{figure}\begin{center}
\epsfig{file=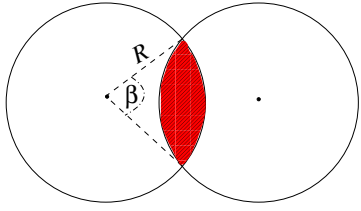, width=5cm}
\end{center}
\caption{\sl Nuclear collision in the transverse plane.}\label{ANGLE}
\end{figure}

To estimate the integrals in the second line of \eq{MAIN} we
need to integrate \eq{I1} over momenta $k_1$
\begin{eqnarray}\label{X3}
&& \int\, d^2 k_1\, d^2 k_2\, P(k_1, k_2, {\un B}) \,
\cos(2(\phi_1-\phi_2))  \nonumber\\ &&  = \, \int_0^\infty \, d k_1 \, k_1 \,  
\frac{N_c \, C_F \, S_\bot}{
k_1^2}\,\frac{4 \, K_2}{\pi^3}\,\int_0^\infty\frac{dz}{z^3}\,\left(
1-e^{-\un z^2 Q_s^2/4}\right)^2\, J_2(k_1z) \approx \frac{N_c C_F
S_\bot Q_s^2 \, K_2 \, \ln 2}{2 \, \pi^3},
\end{eqnarray}
where we assumed that $Q_s^2$ of \eq{SATSCALE} is approximately
independent of transverse coordinate $\un z$ neglecting the logarithm
of \eq{xG}. To complete the calculation of the integrals in \eq{MAIN}
we note that
\beq\label{X4}
\int\, d^2 k_1\, d^2 k_2\, P(k_1, k_2, {\un B}) \, = \, 
\frac{dN}{dy_1}\, \frac{dN}{dy_2}\, = \, \left( c \, 
\frac{S_\bot\, C_F\, Q_s^2}{\as\, 2\, \pi^2} \right)^2.
\eeq
Inserting Eqs. (\ref{I1}), (\ref{I4}), (\ref{X3}) and (\ref{X4}) into
\eq{MAIN} and noting that the integration over rapidities is trivial 
and cancels out between different integrals we obtain
\beq\label{V2-}
v_2(p_T, \un B)\,=\,\as
\,\left(\frac{\pi \, N_c \, K_2}{2 \, \ln 2 \, C_F \, S_\bot \, Q_s^2 \, K_1^2}
\right)^{1/2}\,
\frac{\int_0^\infty\,\frac{dz}{z^3}\, J_2(p_T z)\, \left(
1-e^{-\un z^2 Q_s^2/4}\right)^2} {\int_0^\infty\,\frac{dz}{z^3}\,
J_0(p_T z)\, \left( 1-e^{-\un z^2 Q_s^2/4}\right)^2}.
\eeq
We have changed the transverse momentum back to $p_T$ to comply with
conventional notation. \eq{V2-} gives the minijet contribution to the
differential elliptic flow for collision of two identical nuclei at
given impact parameter $\un B$. $v_2(p_T)$ measured in experiments is
averaged over all impact parameters
\cite{star}
\beq\label{v2p}
v_2(p_T) \, = \, \frac{\int \, d^2 B \, v_2(p_T, \un B) \,
(dN / d^2 p_T \, dy) }{\int \, d^2 B \, (dN / d^2 p_T \, dy)}. 
\eeq
In our model each nucleus is cylindrical, so that the dependence on
the impact parameter $B$ comes only from the nuclear overlap area
$S_\bot(B)$. The overlap area is
\beq\label{SBOT}
S_\bot\,=\,R^2\, (\beta\, - \sin\beta)\,=\,
R^2\,(2\arccos(B/2R)-\sin(2\arccos(B/2R)),
\eeq
where $\beta$ is the opening angle in the transverse plane as shown in
\fig{ANGLE}.
From \eq{DD1} we see that $dN / d^2 p_T \, dy \, \sim \, S_\perp ({\un
B})$ and from \eq{V2-} one concludes that $v_2(p_T, \un B) \, \sim \,
1/\sqrt{S_\perp ({\un B})}$. Therefore after using Eqs. (\ref{DD1})
and (\ref{V2-}) in \eq{v2p} the impact parameter averaging reduces to
\beq
\frac{\int_0^{2R}\,d B\, B\,  \sqrt{2\arccos(B/2R)-\sin(2\arccos(B/2R)}}
{R \, \int_0^{2R}\,d B\, B\, (2\arccos(B/2R)-\sin(2\arccos(B/2R))} \,
\approx \, \frac{.99}{R} \, \approx \, \sqrt{\frac{\pi}{S^A_\perp}},
\eeq
where $S^A_\perp \, = \, \pi \, R^2$ is the cross sectional area of
the nucleus. The final result for $v_2 (p_T)$ reads
\beq\label{V2}
v_2(p_T)\,=\, \as\,\left(\frac{\pi^2 \, N_c \, K_2}{2 \, \ln 2 \, C_F
\, S^A_\perp \, Q_s^2 \, K_1^2} \right)^{1/2}\,
\frac{\int_0^\infty\,\frac{dz}{z^3}\, J_2(p_T z)\,\left(
1-e^{-\un z^2 Q_s^2/4}\right)^2}
{\int_0^\infty\,\frac{dz}{z^3}\, J_0(p_T z) \,\left(
1-e^{-\un z^2 Q_s^2/4}\right)^2}.
\eeq
\eq{V2} is our main result for differential elliptic flow. Let us first 
study the qualitative features of the flow from \eq{V2} before using
it to fit RHIC data in Sect. 4.

In the small momentum region, $p_T \, \ll \, Q_s$, we can neglect the
exponents in the numerator and denominator of \eq{V2}. Expanding the
Bessel functions for small $p_T z$ and cutting off the integrals over
$z$ by $1/p_T$ from above and $1/Q_s$ from below we end up with
\beq\label{small}
v_2(p_T) \, \approx \, \as \,\left(\frac{\pi^2 \, N_c \, K_2}{2 \, \ln
2 \, C_F \, S^A_\perp \, Q_s^2 \, K_1^2} \right)^{1/2}\,
\frac{p_T^2}{4 \, Q_s^2} \, \ln \frac{Q_s}{p_T},\quad\quad p_T \, \ll
\, Q_s.
\eeq
We see that minijet $v_2$ is an increasing function of transverse
momentum, which is in qualitative agreement with the data.

The asymptotic behavior of $v_2$ at high transverse momenta of the
particles, $p_T \, \gg \, Q_s$, can be found by expanding the
exponents in the numerator and denominator of \eq{V2}. Here it is
essential to keep logarithms arising from Eqs. (\ref{SATSCALE}) and
(\ref{xG}). Simple integration yields
\beq\label{large}
v_2(p_T) \, \approx \, \as \,\left(\frac{\pi^2 \, N_c \, K_2}{2 \, \ln
2 \, C_F \, S^A_\perp \, Q_s^2 \, K_1^2} \right)^{1/2}\,
\ln( p_T/\mu),\quad \quad  p_T \, \gg \, Q_s.
\eeq
Therefore we conclude that while the differential elliptic flow given
by \eq{V2} increases for small transverse momenta, at $p_T \, \sim \,
Q_s$ it turns over and its growth slows down to just a logarithmic
increase. As the transverse momentum increases the DGLAP \cite{dglap}
logarithms should become important modifying \eq{large}. The logarithm
in \eq{large} might also be an artifact of our approach and it might
disappear if a more detailed analysis is carried out with inclusion of
evolution effects in the gluon distribution functions. At the moment
we can make an observation that the contribution from minijets to
differential elliptic flow is in qualitative agreement with the data
\cite{starsn}.

To understand the centrality dependence of the elliptic flow coming
from minijets we use the definition \cite{star}
\beq\label{v2bdef}
v_2(B) \, = \, \frac{\int \, d^2 p_T \, dy \, v_2(p_T, \un B) \,
(dN / d^2 p_T \, dy) }{\int \, d^2 p_T \, dy \, (dN / d^2 p_T \, dy)}
\eeq
to obtain using Eqs. (\ref{DD1}) and (\ref{TOTMULT})
\beq\label{v2b}
v_2(B) \, = \, \as \, \left(\frac{2 \, \ln 2 \ \pi \, K_2 \, N_c}{c^2
\, C_F \, S_\perp (B) \, Q_s^2} \right)^{1/2},
\eeq
where $S_\perp (B)$ is now the $B$-dependent nuclear overlap
area. Since $S_\perp (B) \, Q_s^2 \, \sim \, N_{part}$ we may
therefore conclude that minijets flow scales as
\beq\label{bdep}
v_2(B) \, \sim \, \frac{1}{\sqrt{N_{part}}}.
\eeq
Elliptic flow from \eq{bdep} is smaller for central collisions with
large $N_{part}$ and it increases for peripheral collisions with
decreasing $N_{part}$, again in qualitative agreement with the data
\cite{pho,star,phenix}.

\section{Our model versus experimental data}
\renewcommand{\labelenumi}{(\roman{enumi})}

There are three interesting kinematical regions in high energy heavy
ion collisions corresponding to various values of transverse momenta
$p_T$:
\begin{enumerate}
\item  $p_T\lsim Q_s$ where multiple rescatterings of partons dominate. 
 In this region transverse momentum dependence of $v_2$ is given by
 \eq{small}.  Owing to the nonlinear interactions nuclear structure
 functions in region (i) are functions of only one variable $Q/Q_s(x)$
 instead of two $Q$ and $x$. This is often referred to as geometric
 scaling \cite{geom}.
\item  $p_T \gg Q_s$ where the DGLAP \cite{dglap} evolution starts
 to be important. There the leading behavior of $v_2$ is given by
 \eq{large} in which DGLAP equation may introduce logarithmic
 corrections, possibly changing the power of $\ln (p_T/\mu)$. The
 scaling behavior is broken down.
\item At high energies in the  course of quantum evolution 
 the saturation scale $Q_s$ becomes much larger than any soft QCD
 scale $\Lambda$. It was argued both analytically \cite{larry} and
 numerically \cite{geom,satevol,lub} that the geometrical scaling
 holds in a much wider kinematical region than (i). In fact the
 additional scaling region is $Q_s\lsim p_T \lsim Q_s^2/\Lambda$.
 Here even though the scattering amplitude is still far from
 saturation, the nonlinear interactions produce remarkable scaling
 phenomenon. It was argued in \cite{dn} that the saturation scale in
 RHIC kinematical region is of the order of $1\div 1.4$~GeV. Since
 $\Lambda \sim 0.1\div 0.2$~GeV, we expect the scaling to be important
 at $p_T\le 5\div 10$~GeV, i.e., throughout most of the kinematic
 region of the differential elliptic flow data \cite{starsn}.
\end{enumerate}

To compare predictions of our model with data collected by the STAR
collaboration at RHIC \cite{starsn} we should evaluate $v_2$ given by
\eq{V2} and \eq{v2b} at $p_T < 5$~GeV.  This means calculation in the
regions (i) and (iii) where geometric scaling works and the only
relevant dimensional scale is $Q_s$.  Quantum evolution changes the
quasi-classical result of \eq{SATSCALE} so that the saturation scale
acquires energy dependence. Its analytical expression can be
calculated from the nonlinear evolution equation \cite{bk}. There is a
simple way to introduce $Q_s$ into our model preserving geometric
scaling developed by quantum evolution, where $Q_s$ will play a role
of a phenomenological parameter which is used to fit the data. Namely,
by definition the saturation scale is a scale at which the expression
in the Glauber exponent of \eq{WW} equals one
\beq\label{D1}
\frac{1}{4}\,\un z^2\, Q_s^2 (\un z)\,|_{\un
  z^2=4/Q^2_s}\,=\, \frac{Q_{s0}^2}{Q_s^2}\,\ln
\frac{Q_s}{2 \mu}\,=\, 1, 
\eeq
where we have introduced in the quasi-classical (Glauber)
approximation
\beq
Q_{s0}^2 \, = \, \frac{4\pi\,\alpha_s^2\,A}{S_\perp^A},
\eeq
which is an unknown constant for the case of a fully evolved
distribution.  $Q_{s0}$ can be easily expressed in terms of $Q_s$
using \eq{D1}
\beq\label{qs0}
Q_{s0}^2 \, = \, \frac{Q_s^2}{\ln \frac{Q_s}{2 \mu}}.
\eeq 
Substituting \eq{qs0} into \eq{V2} means the following change in the
power of the Glauber exponents
\beq\label{D2}
-\frac{1}{4}\,\un z^2\, Q_{s0}^2\,\ln
\frac{1}{z \mu}  = \,
-\frac{1}{4}\,\un z^2\, Q_s^2 \,
\frac{\ln\frac{1}{z \mu}}{\ln(Q_s/2 \mu)}.
\eeq
Using \eq{D2} in \eq{V2} we can write for the integrals involved
\beq\label{int}
\int_0^\infty\,\frac{dz}{z^3}\, J_n (p_T z)\,\left(
1-e^{- z^2 Q_s^2 (z)/4}\right)^2 \, = \, p_T^2 \,
\int_0^\infty\,\frac{d\xi}{\xi^3}\, J_n (\xi)\,\left( 1 - e^{- (\xi^2 Q_s^2 
/ 4 p_T^2) \, \frac{\ln\frac{2 \, p_T}{\xi
\Lambda}}{\ln(Q_s/\Lambda)}}\right)^2
\eeq
where we defined $\xi \equiv p_T z$, $\Lambda = 2 \mu$ and $n=0 \,
\mbox{or} \, 2$. Geometric scaling \cite{geom} implies that the
distribution functions depend only on one parameter $p_T/Q_s$. To
eliminate the scaling violating terms in \eq{int} we note that the
average value $<z> \, \sim \, 1/<p_T> \, \sim \, 2/Q_s$ and rewrite
the logarithm as
\beq\label{QQ}
\ln\frac{2 \, p_T}{\xi \, \Lambda} \, \approx \, 
\ln\frac{Q_s}{\xi \, \Lambda}
\eeq
where $\Lambda$ would be an independent parameter of our fit
restricted by reasonable possible values of the non-perturbative
scale. The integral of
\eq{int} becomes
\beq\label{int2}
 p_T^2 \,
\int_0^\infty\,\frac{d\xi}{\xi^3}\, J_n (\xi)\,\left( 1 - e^{- (\xi^2 Q_s^2 
/ 4 p_T^2) \, \frac{\ln\frac{Q_s}{\xi \, \Lambda}}{\ln
(Q_s/\Lambda)}}\right)^2.
\eeq
This form of the integrals will be used in \eq{V2} to numerically
estimate $v_2 (p_T)$.

To obtain numerical value of $v_2(p_T)$ we have to estimate the
normalization coefficient $K_1$. To this aim note that the fraction of
the total particle multiplicity per unit of rapidity due to soft
saturation physics in heavy ion collisions at center-of-mass energy
$\sqrt{s}=130$~GeV is given by \cite{dn}
\beq\label{MULTT}   
\frac{dN_{sat}}{dy}\,\approx\, 1.02 \, \frac{3 \, N_\mathrm{part}}{2},
\eeq
where $3/2$ is the conversion factor from charged particles to all
particles multiplicity.  On the other hand, we can calculate the total
particle multiplicity by integrating \eq{DD1} over all momenta $k_1$
which gives
\beq\label{MULYY}
\frac{dN}{dy}\,=\, K_1 \, \frac{2 \, \ln 2\, S_\bot^A \, C_F\, Q_s^2}{\pi^2\,\as}
\, \ln \frac{Q_s}{2 \mu}.
\eeq
Equating \eq{MULTT} and \eq{MULYY} for the head-on collisions ($B=0$)
we obtain
\beq\label{KK}
K_1 \,=\,\frac{3 \, \pi^2 \,\as\, 1.02 \, N_\mathrm{part}(B=0)}{ 4 \,
\ln 2 \, S_\bot^A \, C_F\, Q_s^2\, \ln(Q_s/\Lambda)}.
\eeq
where $N_\mathrm{part}(B=0)=344$ \cite{dn}. For gold nuclei with $\as
= 0.3$, $Q_s= 1.0$~GeV and $\Lambda \approx 0.15$~GeV the
normalization factor is $K_1 \, \approx \, 0.14$.

\begin{figure}
\begin{center}
\epsfig{file=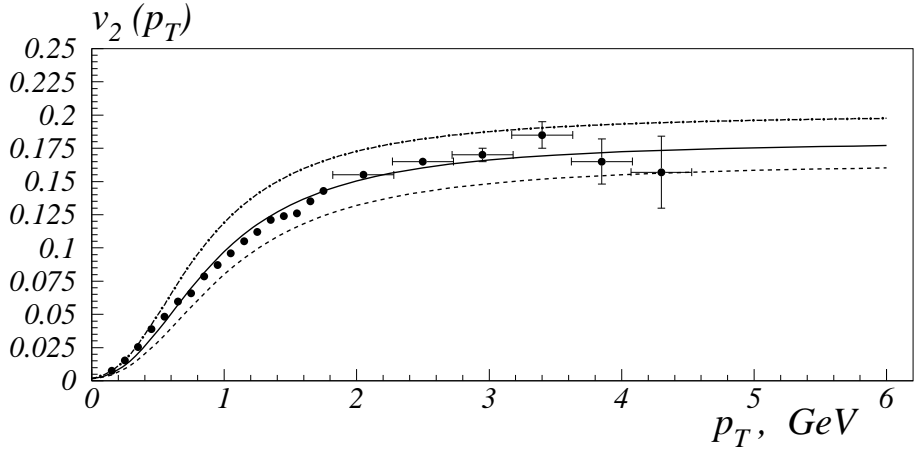, width=13cm}
\end{center}
\caption{\sl Differential elliptic flow data from STAR \cite{starsn} versus 
the predictions of our minijet model. Different lines correspond to
our predictions with different values of the saturation scale: $Q_s=
1.0$~GeV (solid line), $Q_s=1.1$~GeV (dashed line) and $Q_s=0.9$~GeV
(dash-dotted line).  We used the following parameters
$\Lambda$=0.15~GeV, $\as$=0.3, $A$=197 (Gold). }
\label{ptfig}
\end{figure}

With the help of \eq{int2} the integration in \eq{V2} has been done
numerically for $\Lambda \approx 0.15$~GeV. We put $K_2 = 2$ in
\eq{V2} in agreement with collinear factorization approaches 
\cite{coll3,hijing}. $S_\perp^A = \pi R^2$ with the radius given by
Woods--Saxon parameterization $R=1.1 A^{1/3}$~fm. The results are
shown in \fig{ptfig}. It can be seen that
our model describes the high $p_T$ data remarkably well. We are led to
the conclusion that the high momentum tail of the $v_2$ is saturated
by the two-particle correlated minijet production. We can also explain
the turnover of $v_2$ at $p_T\sim 1\div2\sim Q_s$~GeV as being due to
saturation effects in the wave functions of the colliding nuclei at
low $p_T$. Another important observation is that two-particle
correlations in the initial state give significant contribution at low
$p_T$ as one can see in \fig{ptfig}.  Note that the $p^2_T$ increase
of $v_2$ at small momenta may be an artifact of our model which
neglects the change of the anomalous dimension $\gamma$ of the gluon
structure function.  It is well known that correct anomalous dimension
at the boundary of the saturation region is $\gamma\approx 1/2$
\cite{satevol}. It smoothly changes from 0 in the unitarity limit to 1
in the Bjorken limit \cite{satevol}. More accurate analysis is needed
to understand the low $p_T$ behavior of $v_2$ due to two-particle
correlations in minijet production.

Applicability of our model is restricted to the mid-rapidity
kinematical region. However, since the total particle multiplicity is
dominated by the particle production at mid-rapidity the averaging
over all impact parameters in \eq{v2p} gives a reasonable
approximation. The situation is different if we want to apply our
model to describe the centrality dependence of $v_2$. While we expect
it to be in agreement with the data for the most central events, it is
not legitimate to use it for peripheral collisions.

Eqs. (\ref{v2b}) and (\ref{bdep}) imply 
\beq\label{v2bfit}
v_2(B)\,=\,\frac{v_2(B=0)\,\sqrt{N_\mathrm{part}(B=0)}}
{\sqrt{N_\mathrm{part}}}\,=\,
\as \, \left(\frac{\pi \, N_c \, K_2 \,  N_\mathrm{part}(B=0) }
{2 \ln 2 \, C_F \, S_\perp^A \,
Q_s^2 \, N_\mathrm{part}}\right)^{1/2},
\eeq
where we used $c = 2 \ln 2$ \cite{yuaa}.
\begin{figure}
\begin{center}
\epsfig{file=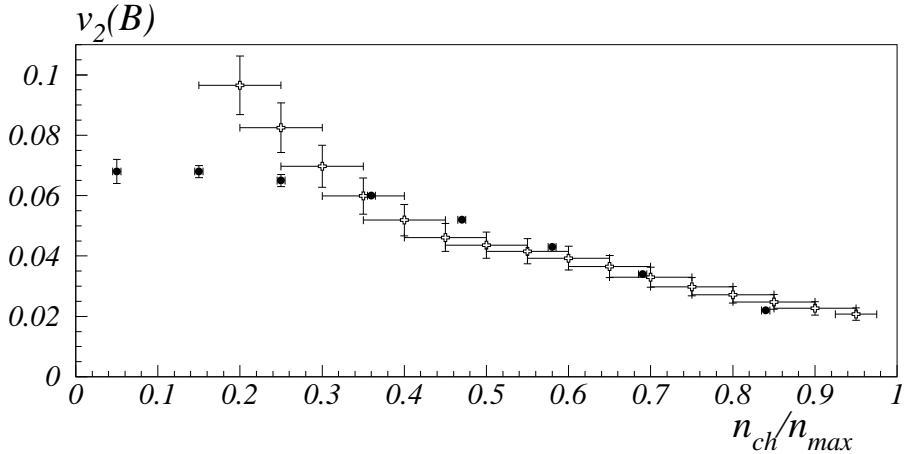, width=13cm}
\end{center}
\caption{\sl Centrality dependence of elliptic flow given by the STAR 
data (black dots) and our fit (empty crosses). We used
$\Lambda$=0.15~GeV, $\as$=0.3, $A$=197, $Q_s$=1~GeV and
$N_\mathrm{part} (B=0)$ = 344 \cite{dn}.}
\label{btfig}
\end{figure}
In \fig{btfig} we show the result of numerical calculations compared
to STAR data \cite{star,starsn}. The data is represented by black dots
while our fit is given by empty crosses. We have used
\eq{v2bfit} with the coefficient $K_2 = 2$. To find the average
$N_\mathrm{part}$ in a given centrality bin we have used the model of
Kharzeev and Nardi \cite{dn}. Horizontal error bars in our fit
correspond to the widths of centrality bins. Vertical error bars
account for our use of cylindrical nuclear shape instead of spherical
and for the uncertainty in $N_\mathrm{part}$ coming from the model of
\cite{dn}.  As expected our model provides a reasonable description
of the data for central events while the fit does not work that well
for peripheral collisions. The saturation scale decreases toward the
large impact parameters where at some point it becomes of the order of
$\Lambda_{QCD}$ and the small coupling approach breaks down. Thus, as
was mentioned above, our approach is not applicable to very peripheral
collisions and can not be expected to work for them. We conclude that
our model gives a reasonable description of $v_2$ centrality
dependence for the values of centrality
$n_\mathrm{ch}/n_\mathrm{max}\ge 0.4$.

The maximum of the centrality distribution of elliptic flow is
probably determined by the onset of non-perturbative effects.  The
flow from \eq{v2b} increases with increasing $B$, until at some point
the saturation scale $Q_s (B)$ becomes of the order of $\Lambda_{QCD}$
and non-perturbative effects take over keeping $v_2 (B)$ approximately
constant for peripheral collisions. (Strictly speaking \eq{v2b} was
derived for cylindrical nuclei with saturation scales independent of
the impact parameter. However in real life $Q_s$ is of course a
function of $B$ \cite{meM,dn,dgn}.) Let us denote the impact parameter
at which the non-perturbative effects take over by $B_0$, so that $Q_s
(B_0) \, \approx \, \Lambda_{QCD}$. Since $Q_s$ is an increasing
function of the center of mass energy $s$ \cite{bk,satevol}, the
impact parameter $B_0$ must also be an increasing function of
$s$. That is as energy increases more and more collisions become
perturbative with the non-perturbative physics being responsible only
for increasingly more peripheral collisions. The maximum of the flow
centrality distribution could be estimated from \eq{v2b} to be $v_2
(B_0) \, \sim \, 1/\sqrt{S_\perp (B_0) \, \Lambda_{QCD}^2}$. With the
increase of energy $B_0$ increases, $S_\perp (B_0)$ decreases and,
therefore, $v_2 (B_0)$ increases, which qualitatively agrees with RHIC
and SPS data \cite{starsn,na49,wa98,pho,star,phenix}.


\section{Conclusions}

In this paper we have calculated the contribution of pairwise
azimuthal correlations in minijet production to elliptic flow
variable. The resulting differential elliptic flow $v_2 (p_T)$ given
by \eq{V2} is in good qualitative (see Eqs. (\ref{small}) and
(\ref{large})) and quantitative (see \fig{ptfig}) agreement with the
emerging RHIC data \cite{pho,star,phenix,starsn}. The centrality
dependence of the minijet contribution to elliptic flow $v_2 (B)$ (see
Eqs. (\ref{v2b}) and (\ref{bdep})) successfully describes the data for
sufficiently central collisions as shown in \fig{btfig}. The maximum
$v_2 (B)$ appears to increase with energy also in agreement with the
data \cite{starsn}.

There are several important questions which still have to be addressed
in the future as more flow data is produced at RHIC. One question
concerns the value of the directed flow $v_1$. It may seem from the
above discussion that using two-particle correlations one might write,
similar to \eq{v2o},
\beq\label{v1}
v_1 (p_T) \, = \, \frac{\left< \, \cos (\phi_1 (p_T) - \phi_2) \,
\right>}{\sqrt{\left< \, \cos (\phi_1 - \phi_2) \, \right>}} 
\eeq
and using the correlations of \eq{DD2} get a non-zero directed
flow. However, let us remind the reader that in the analysis of
directed flow the signs of the weights used in determination of the
reaction plane are reversed in backward hemisphere with respect to the
forward one \cite{elfl,pv}. That is, in the numerator of \eq{v1}, the
contributions from $y_2 > 0$ and $y_2 < 0$ come in with different
signs (we denote the central rapidity of the event by
$y=0$). Therefore, since in our boost-invariant model the two
contributions are identical, the final result for $v_1$ at
mid-rapidity is zero in agreement with the SPS data \cite{na49}. If we
impose some constraint on the rapidity interval of the analyzed
particles making it asymmetric with respect to $y=0$, e.g. analyzing
only particles with $y>0$, the differential directed flow from
minijets would also become non-zero as observed experimentally
\cite{na49}.

In our analysis we have neglected contributions from correlated
production of three and more particles in a single subcollision. The
corresponding diagrams may also contribute to the two-particle
correlation function in \eq{COR}.  Of course they are suppressed by
extra powers of the strong coupling constant $\as$ compared to
\eq{DD2}, though these extra powers are compensated by the large 
logarithms arising due to phase space integration
\cite{dglap,BFKL}. In heavy ion collisions at RHIC the saturation
scale is of the order of $Q_s \, \sim \, 1 \, \mbox{GeV}$
\cite{dn,dgn,yuaa} and the corresponding value of the coupling
constant is $\as (Q_s) \, \approx \, 0.3$, which is not too small and
will also be enhanced by logarithms of energy and transverse
momentum. Thus to have a precise description of the minijet
contribution to elliptic flow one has to resum all these multiple
emission diagrams, which is equivalent to including the effects of
quantum evolution in the double inclusive gluon distribution of
\eq{DD2}. Since at the moment there is even no rigorously derived
expression for double gluon production in AA in the quasi-classical
approximation, including evolution effects in it seems like an
important but not immediate goal. 

To improve the quality of the minijet predictions one has to also
relax the $y_1 \gg y_2$ (or $y_1 \ll y_2$) condition which we imposed
to simplify the calculation. This condition led to absence of rapidity
correlations in the two-particle distribution given by our
\eq{DISTR2}. Also the azimuthal correlations given by \eq{DISTR2} are 
only ``back-to-back'', i.e., the produced particles are correlated
mostly in the opposite directions in the transverse plane. This
behavior seems to contradict recent results reported by PHENIX
\cite{phenix,jr} where the two-particle correlation function $C
(\Delta \phi = \phi_1 - \phi_2)$ has maxima at both $\Delta \phi =
\pi$ (back-to-back) and $\Delta \phi = 0$. Relaxing the $y_1 \gg y_2$
(or $y_1 \ll y_2$) condition would greatly complicate the particle
production calculations \cite{NLO} though would give predictions for
elliptic flow in a much more realistic kinematics. The appropriate
calculations were performed for production of a pair of particles with
$y_1 \sim y_2$ by Leonidov and Ostrovsky in \cite{leostr}. The results
of \cite{leostr} indicate that the two produced particles are indeed
somewhat correlated in rapidity. The analysis of azimuthal
distributions performed in \cite{leostr} also demonstrates enhancement
of both $\Delta \phi = \pi$ and $\Delta \phi = 0$ correlations in the
obtained double particle spectrum, in qualitative agreement with
PHENIX data \cite{phenix,jr}. These correlations should also
contribute to smallness of directed flow $v_1$. A detailed analysis of
elliptic flow employing the full calculation of \cite{leostr} will be
done elsewhere \cite{us}. In a complete analysis one has to put in
realistic geometrical shapes for the nuclei as well. Of course the
effects of different cuts used to analyze the data and various choices
of weights have to be incorporated in the precise analysis too. While
these modifications are likely to introduce some numerical changes in
the fits to the elliptic flow data presented above, we believe that
the qualitative conclusions of importance of the minijet production to
flow variables would remain unchanged.

We therefore conclude that minijets give a large contribution to the
elliptic flow extracted using current flow analysis methods, probably
accounting for most of the flow at large $p_T$ (see
\fig{ptfig}). Differential elliptic flow appears to be sensitive to
saturation physics of the early stages of the collisions. To study QGP
it would be very useful to invent a method of flow analysis that would
be insensitive to minijet contribution and would only measure the
collective effects due to elliptic flow \cite{ollie1,ollie2}. We have
to note, however, that if one tries to calculate the double inclusive
minijet production cross section in the saturation approach
\cite{mv,claa,yuaa} many of the diagrams that would contribute would 
reduce to independent production of two gluons which would then
exchange a gluon with each other. This $2 \rightarrow 2$ process would
constitute the first step in the onset of thermalization
\cite{bmss}. Therefore it appears that correlated production of pairs 
of minijets may be intrinsically related to thermalization. On one
hand this implies that the minijet flow of \eq{V2} is partially due to
collective effects after all. On the other hand it may be impossible
to create an observable which would distinguish these early stage
thermalization effects from the flow of the fully thermalized
quark-gluon plasma.

\section*{Acknowledgements} 

The authors would like to thank Adrian Dumitru, Miklos Gyulassy, Dima
Kharzeev, Roy Lacey, Larry McLerran, Jean-Yves Ollitrault, and Jan Rak
for stimulating and informative discussions. Our thanks go to Derek
Teaney and Raju Venugopalan for pointing out to us a numerical error
in the earlier version of the paper.

The work of Yu. K. was supported in part by the U.S. Department of
Energy under Grant No. DE-FG03-97ER41014 and by the BSF grant $\#$
9800276 with Israeli Science Foundation, founded by the Israeli
Academy of Science and Humanities. The work of K. T. was sponsored in
part by the U.S. Department of Energy under Grant
No. DE-FG03-00ER41132.

\end{document}